\documentclass[aps,pra,preprint,superscriptaddress]{revtex4-1}

\usepackage{amsmath,amssymb,graphicx}
\usepackage[english]{babel}
\usepackage{graphicx}

\begin{document}

\title{Ground state in the finite Dicke model for interacting qubits}

\author{R. A. Robles Robles}
\affiliation{Instituto Nacional de Astrof\'{\i}sica, \'Optica y Electr\'onica, Calle Luis Enrique Erro No. 1, Sta. Ma. Tonantzintla, Pue. CP 72840, M\'exico}

\author{S. A. Chilingaryan}
\affiliation{Departamento de F\'{i}sica, Universidade Federal de Minas Gerais, Caixa Postal 702, 30123-970, Belo Horizonte, MG, Brazil}

\author{B. M. Rodr\'{\i}guez-Lara}
\email{bmlara@inaoep.mx}
\affiliation{Instituto Nacional de Astrof\'{\i}sica, \'Optica y Electr\'onica, Calle Luis Enrique Erro No. 1, Sta. Ma. Tonantzintla, Pue. CP 72840, M\'exico}

\author{Ray-Kuang Lee}
\affiliation{Department of Physics, National Tsing-Hua University, Hsinchu 300, Taiwan.}
\affiliation{ Institute of Photonics Technologies, National Tsing-Hua University, Hsinchu 300, Taiwan.}

\begin{abstract}
We study the ground state of a finite size ensemble of interacting qubits driven by a quantum field.
We find a maximally entangled W-state in the ensemble part of the system for a certain coupling parameters region.
The area of this region decreases as the ensemble size increases and, in the classical limit, becomes the line in parameter space that defines the phase transition of the system.
In the classical limit, we also study the dynamics of the system and its transition from order to disorder for initial energies close to the ground state energy. 
We find that a critical energy providing this transition is related to the minimum of the projection of the total angular momentum of the quantum system in the $z$-direction.
\end{abstract}

%\pacs{03.67.Bg,03.75.Kk,42.50.Pq}

\maketitle
%%%%%%%%%%%%%%%%%%%%%%%%%%%%%%%%%%%%%%%%%%%%%%%%%%%%%%%%%%%%%%%%%%%%%%%%%%%%%%%%%%%%%%%%
\section{Introduction} \label{sec:S1}
%%%%%%%%%%%%%%%%%%%%%%%%%%%%%%%%%%%%%%%%%%%%%%%%%%%%%%%%%%%%%%%%%%%%%%%%%%%%%%%%%%%%%%%

A set of $N_{q}$ interacting qubits coupled to a quantized field may be described by the Hamiltonian
\begin{eqnarray} 
\hat{H} = \omega_{f} \hat{N} + \delta \hat{J}_{z} + \frac{\eta}{N_{q}} \hat{J}_{z}^{2} + \frac{\lambda}{\sqrt{N_{q}}} \left(\hat{a}  + \hat{a}^{\dagger} \right) \hat{J}_{x}, \label{eq:ExtDicke}
\end{eqnarray} 
where the detuning, $\delta = \omega - \omega_{f}$, is defined as the difference between the qubit transition frequency, $\omega$, and the quantized field frequency, $\omega_{f}$, the field-ensemble and the inter-qubit coupling are given by the constants $\lambda$ and $\eta$, in that order, the orbital angular momentum representation, $J_{i}$ with $i=x,y,z,\pm$ such that $\left[\hat{J}_{i},\hat{J}_{j} \right] = i \epsilon_{ijk} \hat{J}_{k}$, describes the qubit ensemble, the  creation (annihilation) operators describe the field, $\hat{a}^{\dagger}$ ($\hat{a}$), and the total number of excitation in the system is given by the operator $\hat{N} = \hat{a}^{\dagger}\hat{a} + \hat{J}_{z} + N_{q}/2$.
This model may be realized with trapped hyperfine ground states of a Bose-Einstein condensate inside a microwave cavity \cite{Chen2006p2,Chen2007p40004,Chen2010p053841,Nagy2008p127} or with arrays of interacting superconducting qubits coupled to the quantum field mode in a coplanar waveguide resonator \cite{Chen2007p055803} where a no-go theorem may \cite{Viehmann2011p113602} or may not \cite{Nataf2010p72} exists, or with coupled nitrogen vacancy centers interacting to a planar microwave cavity \cite{Zou2014p023603}. 

Beyond the fact that this Hamiltonian describes an experimentally feasible system that goes from integrable at $\lambda=0$ \cite{Dukelsky2004p643} to quasi-integrable at $\eta = 0$ \cite{Emary2003p044101}, our interest is twofold.
First, this Hamiltonian is equivalent to the Lipkin-Meshkov-Glick (LMG) model \cite{Lipkin1965p188} in the limit $\lambda = 0$. 
The LMG model produces maximal entanglement at the first (second) order quantum phase transition of the ground state if the coupling is anti-ferromagnetic (ferromagnetic) \cite{Vidal2004p022107,Vidal2004p054101}.  
The ground state phase transition of the Hamiltonian in Eq.\eqref{eq:ExtDicke} has been studied in the thermodynamic limit, $N_{q} \rightarrow \infty$, within and without the rotating wave approximation (RWA) \cite{Chen2007p40004} and in the quantum regime, using coherent states for both the field and ensemble, without the RWA \cite{Chen2010p053841}; these results show the existence of a finite size first order quantum phase transition and a second order super-radiant phase transition. 
Finite size quantum phase transitions in the ground state of the finite Dicke model have been associated with entanglement between the ensemble and the quantum field \cite{Lambert2004p073602,Vidal2006p817} and with bipartite entanglement among qubits due to finite-size effects \cite{Buzek2005p163601,Tsyplyatyev2009p012134,RodriguezLara2010p2443}.
Thus, for a sufficiently low field-ensemble coupling and an adequate inter-qubit coupling, it may be possible to obtain a maximally entangled state of the ensemble in the ground state of the system described by Eq.(\ref{eq:ExtDicke}). 
Entanglement is a fundamental quantum mechanical phenomenon \cite{Schrodinger1935p807,Schrodinger1935p823,Schrodinger1935p844,Einstein1935p777,Bell1964p195} and a precious resource in quantum information processing \cite{Nielsen2000,Preskill2000p127}.
For a qubit ensemble, a $W$-state  \cite{Dur2000p062314} maximizes pairwise entanglement of formation \cite{Koashi2000p050302,Dur2001p020303} and is a robust source of entanglement \cite{Briegel2001p910,Sen(De)2003p062306}; i.e., it retains maximal bipartite quantum correlations whenever any pair of qubits are traced out. 
We will show that such a state is produced in the ground state of the Hamiltonian in Eq.\eqref{eq:ExtDicke} for a given parameter regime.
Second, it has been recently shown that the finite size Dicke model, $\eta=0$, shows two excited-state quantum phase transitions; one for any given coupling at an energy rate $2 E/(\omega N_{q}) = 1$  \cite{PerezFernandez2011p033802,PerezFernandez2011p046208} and another at the superradiant phase at an energy rate  $2 E/(\omega N_{q}) = -1$ \cite{BastarracheaMagnani2014p032101}. 
These transitions has been shown as peaks in the Peres lattice \cite{Peres1984p1711} of the system and a transition from order to disorder in the equivalent classical system has been shown around these energy rates \cite{BastarracheaMagnani2014p032102}.
The latter, the so-called excited state quantum phase transition, is related to the unstable fixed points of the classical Dicke Hamiltonian. 
Some of us have shown the existence of stable and unstable fixed points that produce symmetric and asymmetric dynamics in the classical equivalent of Hamiltonian \eqref{eq:ExtDicke} under the RWA \cite{RodriguezLara2011p016225}.

In the following, within the RWA and for weak inter-qubit coupling, we will show analytically that the maximal entanglement produced by the quantum phase transition in the Lipkin-Meshkov-Glick model is retained in the Dicke model for interacting qubits.
Furthermore, for a fixed field-ensemble coupling, $\lambda$, the inter-qubit coupling, $\eta$, defines a series of first order phase transitions related to the number of qubits in the ensemble, $N_{q}$, while, for a fixed inter-qubit coupling, there exists one second order phase transition related to the couplings ratio \cite{Chen2010p053841} and a series of first order transitions similar to those in the Dicke model under the RWA \cite{Buzek2005p163601,Tsyplyatyev2009p012134}. 
Also, we will show analitically in the weak coupling regime, and numerically in general, that the inclusion of counter-rotating terms does not erase the possibility of obtaining a maximally entangled multi-qubit state.
Then, in the third section, we will find the fixed points of the classical analog of the Dicke model for interacting qubits without the RWA, and calculate the critical coupling parameter related to them; this parameter should be identical to that of the quantum case with an ensemble of infinite size.
Also, we will show that a transition from order to disorder in the classical dynamics appears at exactly the value of the scaled energy corresponding to the minimum of the $\hat{J}_{z}$ operator in the quantum system; in our case this scaled energy is $2E/(\omega N_{q}) \le -1$ and depends linearly on the size of the qubit ensemble.

%%%%%%%%%%%%%%%%%%%%%%%%%%%%%%%%%%%%%%%%%%%%%%%%%%%%%%%%%%%%%%%%%%%%%%%%%%%%%%%%%%%%%%%%
\section{Entanglement in the ground state}
%%%%%%%%%%%%%%%%%%%%%%%%%%%%%%%%%%%%%%%%%%%%%%%%%%%%%%%%%%%%%%%%%%%%%%%%%%%%%%%%%%%%%%%%

Here, we want to show that it is possible to find a maximally entangled W-state in the ground state of our Hamiltonian model for a certain parameter set. 
For this reason, we need to start with the exact ground state in the RWA.

\subsection{Rotating wave approximation}
The Hamiltonian in Eq.(\ref{eq:ExtDicke}) under the RWA,
\begin{eqnarray} 
\hat{H}_{RWA} = \omega_{f} \hat{N} + \delta \hat{J}_{z} + \frac{\eta}{N_{q}} \hat{J}_{z}^{2} + \frac{\lambda}{2 \sqrt{N_{q}}} \left(\hat{a} \hat{J}_{+}  + \hat{a}^{\dagger} \hat{J}_{-} \right), \label{eq:ExtTavisCumings}
\end{eqnarray}  
conserves the total number of excitations, thus the corresponding Hilbert space can be partitioned into subspaces where the Hamiltonian becomes a square matrix  $H^{(n)} = H_{O}^{(n)} +  H_{I}^{(n)}$ with 
\begin{eqnarray}
H_{O}^{(n)} &=&  \omega_{f} \left( n - \frac{N_{q}}{2} \right) \mathbb{I}_{\tilde{n}},\\ 
\left(  H_{I}^{(n)} \right)_{i,j} &=& \delta_{i,j} d_{i} + \frac{\lambda}{2 \sqrt{ N_{q}}} \left( \delta_{i,j-1} o_{j}  + \delta_{i,j+1} o_{j+1}  \right),
\end{eqnarray}
where the identity matrix of rank $d$ is given by $\mathbb{I}_{d}$, the row and column labels are in the range $i,j = \tilde{n},\tilde{n}+1, \ldots,n,$ where $\tilde{n} = \mathrm{max}(0,n-N_{q})$, for the photon number $n=0,1,2,\ldots$ The symbol $\delta_{i,j}$ stands for Kronecker delta and the diagonal and off-diagonal terms are defined as,
\begin{eqnarray}
d_{j} & = & \left(n-j-\frac{N_{q}}{2} \right) \left[ \delta +  \frac{\eta}{N_{q}} \left(n-j-\frac{N_{q}}{2}\right) \right], \\
o_{j} & = & \sqrt{j (N_{q}+j-n)(n-j+1) }.
\end{eqnarray}
In each subspace the square matrix is a tri-diagonal symmetric real matrix with positive off-diagonal terms, \textit{i.e.,} a Jacobi matrix, and its eigenvalues can be found analytically \cite{Swalen1961p736,Pierce1961p740,Haydock1972p2845,Yamani1997p2889}. 
The ground state of the system is found as the lowest eigenvalue for the set $\{ H^{(n)} \} $ for all values of $n$.
Furthermore, a first order quantum phase transition is located at the intersection of two ground state energies belonging to contiguous subspaces.

The first ground state structure, which we will call vacuum phase from now on, corresponds to the vacuum field and the qubit ensemble state with zero excitation,
\begin{eqnarray}
|\psi_{g}^{(0)}\rangle = |0\rangle \left\vert \frac{N_{q}}{2},-\frac{N_{q}}{2} \right\rangle.
\end{eqnarray} 
A first quantum phase transition, in a series of first order quantum phase transitions, is found at the critical coupling strength, 
\begin{equation} \label{eq:CritStr}
\lambda_{c} =  2 \sqrt{ \left[ \omega + \left(\frac{1}{N_{q}} -1 \right) \eta \right] \omega_{f} }, 
\end{equation}
with $0 \le \eta \le   N_{q} \omega /(N_{q} -1)$. 

After this critical curve in parameter space, the ground state becomes,   
\begin{eqnarray} \label{eq:GroundStateOneExc}
|\psi_{g}^{(1)} \rangle &=& c_{0}^{(1)} |0\rangle \left\vert \frac{N_{q}}{2}, 1- \frac{N_{q}}{2} \right\rangle   + c_{1}^{(1)} |1\rangle  \left\vert \frac{N_{q}}{2}, - \frac{N_{q}}{2} \right\rangle. 
\end{eqnarray}
The amplitudes are given by the expressions $c_{0}^{(1)} = h/(h^2+1)^{1/2}$ and $c_{1}^{(1)} = 1/(h^2+1)^{1/2}$ related to the amplitude parameter,
\begin{eqnarray} \label{eq:AmpParam}
h =  \frac{ ( 1 -1/ N_{a} ) \eta + \delta - \{4 \lambda^{2} + \left[ (1 -1/N_{a}) \eta - \delta \right]^{2}\}^{1/2}}{2 \lambda}.
\end{eqnarray}
In this first non-vacuum phase, the ground state is fully separable, $|\psi_{g}^{(1)} \rangle = |1\rangle  \left\vert N_{q}/2 , - N_{q}/2 \right\rangle$, in the limit $h \rightarrow 0$ that occurs when the field-ensemble coupling, $\lambda$, is dominant.
A second type of ground state occurs in the limit $h \rightarrow 1$, where there is maximal entanglement between the field and the qubit ensemble, $|\psi_{g}^{(1)} \rangle = \left( |0\rangle \left\vert N_{q}/2, 1- N_{q}/2 \right\rangle   +  |1\rangle  \left\vert N_{q}/2, - N_{q}/2 \right\rangle \right) / \sqrt{2}$. 
And a third type where there ground state of the whole system is separable, $|\psi_{g}^{(1)} \rangle = |0\rangle \left \vert N_{q}/2, 1- N_{q}/2 \right\rangle$,  occurs in the limit $h \rightarrow \infty$, in other words when $\lambda$ is small compared to the denominator of Eq.\eqref{eq:AmpParam}.
Here, the ensemble part is maximally entangled as the qubit ensemble state  $\left\vert N_{q}/2, 1- N_{q}/2 \right\rangle$ is a W-state.
Note that the transition from one case to the other is continuous in this first non-vacuum ground state and any extended Dicke model that conserves the total number of excitations has a first non-vacuum phase of this form; e.g., a Dicke model that includes a quantized field non-linearity \cite{RodriguezLara2010p2443}.
Also, the following phase of the ground state, 
\begin{eqnarray}
|\psi_{g}^{(2)} \rangle &=& \sum_{j=0}^{2} c_{0}^{(j)} |j\rangle \left\vert \frac{N_{q}}{2}, 2-j- \frac{N_{q}}{2} \right\rangle, 
\end{eqnarray}
will make the area for the first non-vacuum ground state smaller as the number of qubits in the ensemble grows.
This may be a problem in the BEC realization where the size of the ensemble is large but this is not a problem in a circuit-QED realization where the number of qubits is small.

%%%%%%%%%%%%%%%%%%%%%%%%%%%%%%%%%%%%%%%%%%%%%%%%%%%%%%%%%%%%%%%%%%%%%%%%%%%%%%%%%%%%%%%
\subsection{Full model under weak couplings}
%%%%%%%%%%%%%%%%%%%%%%%%%%%%%%%%%%%%%%%%%%%%%%%%%%%%%%%%%%%%%%%%%%%%%%%%%%%%%%%%%%%%%%%

Now, we want to show that the maximally entangled $W$-state shown at the first non-vacuum phase survives the inclusion of counter-rotating terms.
We will use the unitary transformation,
\begin{equation}
\hat{U} = e^{- \imath \xi (\hat{a} + \hat{a}^{\dagger} ) \hat{J}_{y} } , \quad \xi =  \lambda N_{q}^{-1/2} / (\omega + \omega_{f}),
\end{equation}
in the weak coupling regime, $\lambda \ll \omega$ such that $\xi \ll 1$.
We will also require a weak intra-ensemble interaction, $\eta \ll \omega$, then it is possible to approximate, 
\begin{eqnarray}
\tilde{H}_{CR} &=& \hat{U}^{-1} \hat{H}_{CR} \hat{U}, \nonumber \\
& \approx & \omega_{f}  \hat{a}^{\dagger} \hat{a} +  \omega  \hat{J}_{z} + \frac{\eta}{N_{q}} \hat{J}_{z}^{2}  + \frac{\tilde{\lambda}}{\sqrt{N_{q}}} (\hat{a} \hat{J}_{+} + \hat{a}^{\dagger} \hat{J}_{-}), 
\end{eqnarray}
with an auxiliary coupling strength $\tilde{\lambda} = 2 \omega_{f} \lambda / (\omega + \omega_{f})$. 
In other words, we have made an effective rotating wave approximation, and we already know that such a system has a first order quantum phase transition at the critical value,
\begin{eqnarray} 
\lambda_{c}^{(CR)} = (\omega + \omega_{f}) \sqrt{\frac{1}{\omega_{f}}\left[ \omega + \left(\frac{1}{N_{q}} -1 \right) \eta \right]}.  \label{eq:CritStrWeak}
\end{eqnarray}
Note that on resonance the expression for the first critical coupling in the weak coupling regime, Eq.(\ref{eq:CritStrWeak}), is  equal to the critical coupling in the rotating wave approximation, Eq.(\ref{eq:CritStr}). 
The ground state at the first non-vacuum phase is described again by Eq.(\ref{eq:GroundStateOneExc}) if we make the substitution $\lambda \rightarrow \tilde{\lambda}$. 
Then, the maximally entangled $W$-state survives the inclusion of counter-rotating terms for the region of interest; \textit{i.e.}, the weak coupling regime.

%%%%%%%%%%%%%%%%%%%%%%%%%%%%%%%%%%%%%%%%%%%%%%%%%%%%%%%%%%%%%%%%%%%%%%%%%%%%%%%%%%%%%%%
\subsection{Semiclassical analysis}
%%%%%%%%%%%%%%%%%%%%%%%%%%%%%%%%%%%%%%%%%%%%%%%%%%%%%%%%%%%%%%%%%%%%%%%%%%%%%%%%%%%%%%%

Here we present a semiclassical analysis of the ground state just for the sake of completeness.
In the Holstein-Primakoff representation of SU(2) \cite{Katriel1986p2332} for a large number of qubits in the ensemble, $\hat{J}_{z} = \hat{b}^{\dagger} \hat{b} - N_{q}/2$, $\hat{J}_{+} \approx \sqrt{N_{q}} \hat{b}^{\dagger} \left( 1 - \hat{b}^{\dagger} \hat{b}/ (2 N_{q}) \right) $ and $\hat{J}_{-} \approx \sqrt{N_{q}}  \left( 1 - \hat{b}^{\dagger} \hat{b}/ (2 N_{q}) \right)~\hat{b}$, the Hamiltonian in Eq.\eqref{eq:ExtDicke}, up to a constant, reduces to
\begin{eqnarray}
\hat{H} &\approx& \omega_{f} \hat{a}^{\dagger} \hat{a} + \left( \omega - \eta \right) \hat{b}^{\dagger} \hat{b} + \frac{\eta}{N_{q}} \left( \hat{b}^{\dagger} \hat{b}\right)^{2} + \nonumber \\
&& \frac{\lambda}{2} \left( \hat{a}^{\dagger} + \hat{a} \right) \left[ \hat{b}^{\dagger} \left( 1- \frac{\hat{b}^{\dagger} \hat{b}}{N_{q}} \right) + \left( 1- \frac{\hat{b}^{\dagger} \hat{b}}{N_{q}} \right) \hat{b} \right].
\end{eqnarray}
Thus, we can consider a coherent state for both the field and the qubit ensemble, $\vert \alpha \rangle_{f} \vert \beta \rangle_{q}$, to calculate the mean energy, up to a constant,
\begin{eqnarray}
\langle \hat{H} \rangle &\approx& \omega \vert \alpha \vert^2 + \left[ \omega - \eta \left( 1 - \frac{\vert\beta \vert^{2} - 1}{N_{q}} \right) \right] \vert \beta \vert^{2} \nonumber \\
&& +\frac{\lambda}{2} \left( 1- \frac{\vert \beta \vert^{2}}{2N_{q}} \right)\left( \alpha + \alpha^{\ast} \right) \left( \beta + \beta^{\ast} \right).
\end{eqnarray}
In order to find the inflection points for this semiclassical energy, we derive with respect to the real and imaginary parts of both $\alpha$ and $\beta$ and solve the system $\partial \langle \hat{H} \rangle / \partial x = 0$ with $x=\alpha_{R}, \alpha_{I}, \beta_{R}, \beta_{I}$.
The trivial solution is the ground state $\vert 0 \rangle_{f} \vert 0 \rangle_{q}$ and we check for intersections with the remaining six solutions, two of them do not intersect the ground state and the remaining four do it at the semiclassical critical coupling parameter, 
\begin{eqnarray}
\lambda_{c}^{(SC)} &=& \sqrt{ \left[ \omega + \left(  \frac{1}{N_{q}} -1 \right)\eta \right] \omega_{f}}, \\
&=& \frac{\lambda_{c}}{2}
\end{eqnarray}
 at which a second order superradiant phase transition occurs \cite{Chen2007p40004, Chen2010p053841}. 
 It is half the critical strength found for the case without counter-rotating terms, $\lambda_{c}$ in \eqref{eq:CritStr}, as expected from what happens for the Dicke model in the classical limit result, where accounting for counter-rotating terms halves the critical coupling found without the counter-rotating terms \cite{Carmichael1973p47}.

%%%%%%%%%%%%%%%%%%%%%%%%%%%%%%%%%%%%%%%%%%%%%%%%%%%%%%%%%%%%%%%%%%%%%%%%%%%%%%%%%%%%%%%
\subsection{Numerical analysis}
%%%%%%%%%%%%%%%%%%%%%%%%%%%%%%%%%%%%%%%%%%%%%%%%%%%%%%%%%%%%%%%%%%%%%%%%%%%%%%%%%%%%%%%

\begin{figure}
\centering \includegraphics[scale= 1]{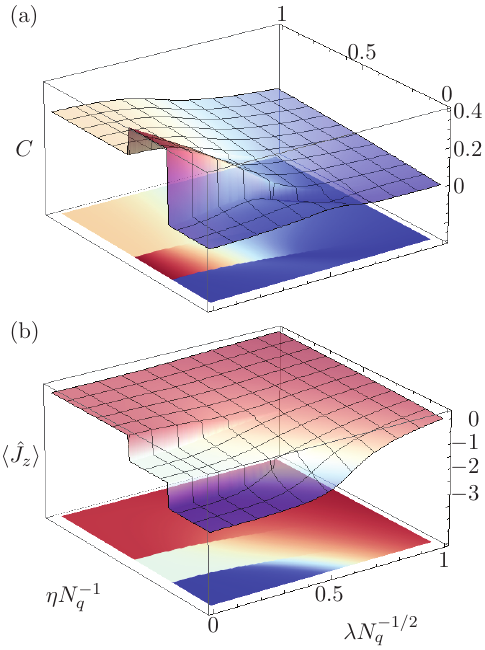}
\caption{(Color online). (a) Entangled web concurrence and (b) mean value of $\hat{J}_{z}$ for the ground state of the model Hamiltonian in \eqref{eq:ExtDicke} with five qubits, $N_{q}=5$, under resonant interaction with the quantized field, $\omega=\omega_{f}$. The qubit interaction, $\eta$, and field-ensemble coupling, $\lambda$, are given in units of the field frequency, $\omega_{f}$.} \label{fig:Figure1}
\end{figure}

\begin{figure}
\centering \includegraphics[scale= 1]{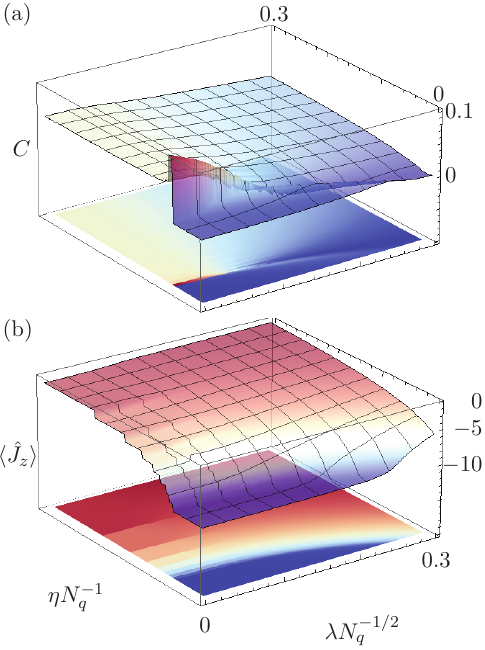}
\caption{(Color online). (a) Entangled web concurrence and (b) mean value of $\hat{J}_{z}$ for the ground state of the model Hamiltonian in \eqref{eq:ExtDicke} with twenty qubits, $N_{q}=20$, under resonant interaction with the quantized field, $\omega=\omega_{f}$. The qubit interaction, $\eta$, and field-ensemble coupling, $\lambda$, are given in units of the field frequency, $\omega_{f}$.} \label{fig:Figure2}
\end{figure}

In order to find the numerical ground state of the Hamiltonian in Eq.\eqref{eq:ExtDicke} we will follow a coherent state approach \cite{Chen2008p051801}. 
For this reason we will move to the frame defined by, 
\begin{eqnarray}
\vert \psi \rangle = \hat{R}_{y}\left( \frac{\pi}{2} \right)  \hat{D}\left(\frac{\tilde{\lambda}}{\omega_{f}} \hat{J}_{z}\right) \vert \xi \rangle, 
\end{eqnarray}
with the rotation around $\hat{J}_{y}$ given as $\hat{R}_{y}\left(\varphi \right) = e^{i \varphi \hat{J}_{y}}$ and the displacement operator in the form $\hat{D}\left(\beta\right)= e^{\beta \hat{a}^{\dagger} - \beta^{\ast} \hat{a}}$ and the size-scaled coupling defined as $\tilde{\lambda} = \lambda / \sqrt{N_{q}}$.
In the new frame, the system is ruled by an effective Hamiltonian, 
\begin{eqnarray}
\hat{H}_{D} &=& \hat{D}^{\dagger} \left(\frac{\tilde{\lambda}}{\omega_{f}} \hat{J}_{z}\right ) \hat{R}^{\dagger}_{y}\left( \frac{\pi}{2} \right) \hat{H} \hat{R}_{y}\left( \frac{\pi}{2} \right)  \hat{D}\left(\frac{\tilde{\lambda}}{\omega_{f}} \hat{J}_{z}\right),  \\
&=& \omega_{f} \hat{a}^{\dagger}\hat{a} - \frac{\tilde{\lambda}^{2}}{ \omega _{f} } \hat{J}_{z}^{2} + \frac{\omega}{2} \left[ \hat{J}_{+} \hat{D}^{\dagger}\left(\frac{\tilde{\lambda}}{\omega_{f}}\right) + \hat{J}_{-} \hat{D}\left(\frac{\tilde{\lambda}}{\omega_{f}}\right)\right]  \nonumber \\
&& +\frac{\eta}{4 N_{q}} \left[ \hat{J}_{+} \hat{D}^{\dagger}\left(\frac{\tilde{\lambda}}{\omega_{f}}\right) + \hat{J}_{-} \hat{D}\left(\frac{\tilde{\lambda}}{\omega_{f}}\right)\right]^{2},
\end{eqnarray}
that is amenable to numerical diagonalization.
In the following we show some numerical results that confirm our prediction of a maximally entangled qubit ensemble in the ground state of the model. 
Figure \ref{fig:Figure1}(a) and Fig. \ref{fig:Figure2}(a) show the concurrence, $C$, for the entangled web \cite{Koashi2000p050302} of qubit ensembles with $N_{q}=5$ and $N_{q}=20$, in that order.
A maximum is reached in the first non-vacuum phase of the ground state and its value is the expected entangled web concurrence maximum of $2/N_{q}$.
The critical coupling strength found in the RWA under weak coupling limit and in the semi-classical limit are in good agreement with the numerical results.
Figure \ref{fig:Figure1}(b) and Fig. \ref{fig:Figure2}(b) show the mean value of $\hat{J}_{z}$.
Note that a single excitation is present for the parameter region where maximum entanglement between the qubits is found as expected \cite{Koashi2000p050302,Dur2001p020303}.

%%%%%%%%%%%%%%%%%%%%%%%%%%%%%%%%%%%%%%%%%%%%%%%%%%%%%%%%%%%%%%%%%%%%%%%%%%%%%%%%%%%%%%%%
%%%%%%%%%%%%%%%%%%%%%%%%%%%%%%%%%%%%%%%%%%%%%%%%%%%%%%%%%%%%%%%%%%%%%%%%%%%%%%%%%%%%%%%%

%%%%%%%%%%%%%%%%%%%%%%%%%%%%%%%%%%%%%%%%%%%%%%%%%%%%%%%%%%%%%%%%%%%%%%%%%%%%%%%%%%%%%%%%
\section{Classical analog}
%%%%%%%%%%%%%%%%%%%%%%%%%%%%%%%%%%%%%%%%%%%%%%%%%%%%%%%%%%%%%%%%%%%%%%%%%%%%%%%%%%%%%%%%
We are also interested in the phase transition of our system related to the classical limit; i.e. making the substitution $N_{q} \rightarrow \infty$ in the critical semi-classical coupling $\lambda_{c}^{(SC)}$.
Some of us have found a relation between the critical value in the semi-classical limit and the transition for fixed points of the classical analog of interacting qubits driven by a quantum field under the RWA \cite{RodriguezLara2011p016225}.
For this reason, we will explore the classical analog of Hamiltonian \eqref{eq:ExtDicke}, which may be obtained by substituting the ensemble operators by the classical angular momentum canonical pair \cite{Aguiar1992p291}, $\left\{ j_{z}, \phi \right\}$, and the field operators by the classical harmonic oscillator canonical pair,  $\left\{ p, q \right\}$,
\begin{eqnarray}
\hat{a} &\longrightarrow& \frac{1}{\sqrt{2}} \left( q + i p\right), \\
\hat{a}^{\dagger} &\longrightarrow& \frac{1}{\sqrt{2}} \left( q - i p\right), \\
\hat{J_{z}} &\longrightarrow& j_{z}, \\
\hat{J_{x}} &\longrightarrow& \sqrt{ j^2-j_{z}^2} ~ \cos\phi, \\
\hat{J_{y}} &\longrightarrow& \sqrt{ j^2-j_{z}^2} ~ \sin\phi,
\end{eqnarray}
where the variable $\phi$ is the azimuthal angle of the Casimir vector $\vec{j}= (j_{x},j_{y},j_{x})$ of constant magnitude $j= N_{q} / 2$. 
Thus, the equivalent classical Hamiltonian,
\begin{eqnarray}
H &=& \frac{\omega_{f}}{2} (q^2+p^2)+\omega j_{z}+\frac{\eta}{2 j} j_{z}^2 + q \lambda \sqrt{j} \sqrt{1 -\frac{j_{z}^2}{j^2}} ~ \cos \phi, \nonumber \\  
\end{eqnarray}
delivers the following equations of motion,
\begin{eqnarray}
\frac{dq}{dt}&=& \omega_{f} p, \\
\frac{dp}{dt}&=& - \omega_{f} q -  \lambda \sqrt{j} \sqrt{1 -\frac{j_{z}^2}{j^2}} ~\cos \phi, \\
\frac{d \phi}{dt}&=& \omega +  \frac{j_{z}}{j} \left[ \eta - \frac{ q \lambda \cos \phi}{\sqrt{j}  \sqrt{1 -\frac{j_{z}^2}{j^2}}}  \right], \\
\frac{dj_{z}}{dt}&=& q \lambda \sqrt{j} \sqrt{1 -\frac{j_{z}^2}{j^2}} ~\sin \phi. 
\end{eqnarray}

%%%%%%%%%%%%%%%%%%%%%%%%%%%%%%%%%%%%%%%%%%%%%%%%%%%%%%%%%%%%%%%%%%%%%%%%%%%%%%%%%%%%%%%%
\subsection{Stable fixed points}
\begin{figure}
\centering \includegraphics[scale= 1]{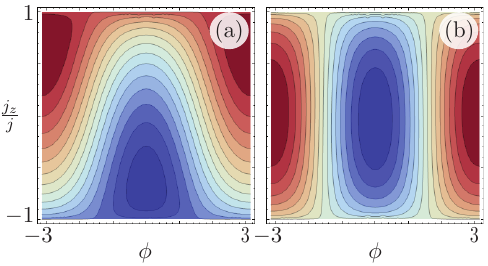}
\caption{(Color online). Energy landscape near the fixed points, $q = q^{(s)}$ and $p=0$, on resonance, $\omega = \omega_{f}$, and fixed inter-qubit coupling, $\eta = 0.1 \omega_{f}$, for ensemble-field couplings (a) $\lambda = 1.25 \lambda_{c}^{(C)}$ and (b)  $\lambda = 6 \lambda_{c}^{(C)}$.} \label{fig:Figure3}
\end{figure}

%%%%%%%%%%%%%%%%%%%%%%%%%%%%%%%%%%%%%%%%%%%%%%%%%%%%%%%%%%%%%%%%%%%%%%%%%%%%%%%%%%%%%%%%
We know from the quantum analysis that the vacuum field and the ensemble without excitation,
\begin{eqnarray}
\left\{ q, p, j_{z},\phi\right\} = \left\{ 0, 0,-j, \phi \right\},
\end{eqnarray}  
is the ground state of the system with energy, 
\begin{eqnarray}
E\left(0, 0,-j, \phi \right) = - j \left( \omega - \frac{\eta}{2} \right).
\end{eqnarray}
Note that this set of variables is not a fixed point of the equations of motion.
Now, our classical mechanics system has a different distribution of fixed points compared to non-interacting qubits \cite{BastarracheaMagnani2014p032101}.
In our case, a set of stable fixed points are given by the following parameters,
\begin{eqnarray}
\left\{ q, p, j_{z},\phi\right\} = \left\{ - q_{(s)}, 0, j_{z}^{(s)}, 0 \right\}, \left\{ q_{(s)}, 0, j_{z}^{(s)}, \pi \right\},
\end{eqnarray} 
with auxiliary definitions,
\begin{eqnarray}
q_{(s)} &=& \frac{\lambda}{\omega_{f}} \sqrt{\frac{j^2 - (j_{z}^{(s)})^{2}}{j}},\\
j_{z}^{(s)} &=& -\frac{j \omega \omega_{f}}{\lambda^{2} +  \eta \omega_{f}},
\end{eqnarray}
and energy,
\begin{eqnarray}
E\left( q_{(s)}, 0, j_{z}^{(s)}, 0 \right) = - \frac{j}{2} \left[ \frac{\lambda^{2}}{\omega_{f}} + \frac{\omega^{2} \omega_{f}}{ \left( \lambda^{2} + \omega_{f} \eta \right)} \right].
\end{eqnarray}
This energy and that for the set related to the quantum ground state before the phase transition intersect at the critical coupling, 
\begin{eqnarray}
\lambda_{c}^{(C)} &=&  \sqrt{ \left( \omega  -\eta \right) \omega_{f}}, \\
&=& \left. \frac{\lambda_{c}}{2} \right\vert_{N_{q} \rightarrow \infty},
\end{eqnarray}
i.e., it is just the semi-classical coupling found before for an ensemble of infinite size.
Thus, we obtain the expected value for the critical coupling.
Figure \ref{fig:Figure3} shows the value of the Hamiltonian,
\begin{eqnarray}
H(j_{z},\phi) &=&  \omega j_{z} + \frac{\eta}{2j} j_{z}^{2} + \frac{\omega_{f}}{2} q_{(s)}^{2}  + \lambda q_{(s)} \sqrt{j} \sqrt{1- \frac{j_{z}}{j}} ~\cos\phi, \nonumber \\
\end{eqnarray}
for stable fixed points parameters,
\begin{eqnarray}
\{ q,p \}= \left\{ q_{(s)},0  \right\},
\end{eqnarray}
for couplings above the critical coupling.
It is straightforward to see that a global minimum is located in the south pole of the sphere, $\left\{ j, \theta, \phi \right\}$, and moves towards the equator after crossing the critical value $\lambda_{c}^{(C)}$.
Near the fixed points, it is possible to find stable periodical oscillations.
These Rabi oscillations localize in a section of the available phase space for values close to the critical coupling; e.g., Fig. \ref{fig:Figure3}(a).  
Note that here the stable fixed points are simpler than those under the RWA \cite{RodriguezLara2011p016225} where one could immediately identify both a Rabi and Josephson regime.
%%%%%%%%%%%%%%%%%%%%%%%%%%%%%%%%%%%%%%%%%%%%%%%%%%%%%%%%%%%%%%%%%%%%%%%%%%%%%%%%%%%%%%%%
%%%%%%%%%%%%%%%%%%%%%%%%%%%%%%%%%%%%%%%%%%%%%%%%%%%%%%%%%%%%%%%%%%%%%%%%%%%%%%%%%%%%%%%%

\begin{figure}
\centering \includegraphics[scale= 1]{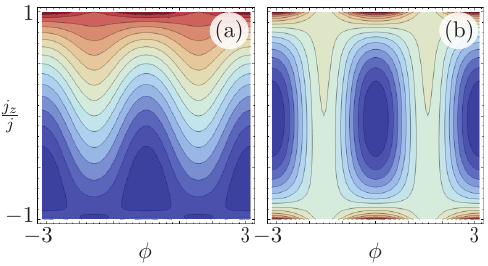}
\caption{(Color online). Energy landscape for parameter values, $q = q^{(s)} \cos \phi$ and $p=0$, on resonance, $\omega = \omega_{f}$, and fixed inter-qubit coupling, $\eta = 0.1 \omega_{f}$, with ensemble-field couplings (a) $\lambda =1.25 \lambda_{c}^{(C)}$ and (b)  $\lambda = 6 \lambda_{c}^{(C)}$.} \label{fig:Figure4}
\end{figure}

So far, it has been shown that the classical Hamiltonian provides a landscape where stable Rabi oscillations may appear. 
In order to explore the energy landscape far from the fixed point described above, we may choose the parameter values,
\begin{eqnarray}
\{ q,p \}= \left\{ q_{(s)} \cos \phi , 0 \right\}.
\end{eqnarray}
Such an approach allow us to see both global minima related to the two fixed points of the system, as shown in Fig. \ref{fig:Figure4}, but provides us with no further information. 
A rigorous analysis of the classical model may be of interest to study both stable and unstable fixed points, as it was done for the model under the RWA \cite{RodriguezLara2011p016225}, but, at the moment, we will just focus on sampling initial conditions and their evolution in phase space near the energy minima.

%%%%%%%%%%%%%%%%%%%%%%%%%%%%%%%%%%%%%%%%%%%%%%%%%%%%%%%%%%%%%%%%%%%%%%%%%%%%%%%%%%%%%%%%
%%%%%%%%%%%%%%%%%%%%%%%%%%%%%%%%%%%%%%%%%%%%%%%%%%%%%%%%%%%%%%%%%%%%%%%%%%%%%%%%%%%%%%%%

%%%%%%%%%%%%%%%%%%%%%%%%%%%%%%%%%%%%%%%%%%%%%%%%%%%%%%%%%%%%%%%%%%%%%%%%%%%%%%%%%%%%%%%%
\subsection{Order and disorder}
%%%%%%%%%%%%%%%%%%%%%%%%%%%%%%%%%%%%%%%%%%%%%%%%%%%%%%%%%%%%%%%%%%%%%%%%%%%%%%%%%%%%%%%%
%%%%%%%%%%%%%%%%%%%%%%%%%%%%%%%%%%%%%%%%%%%%%%%%%%%%%%%%%%%%%%%%%%%%%%%%%%%%%%%%%%%%%%%%
\begin{figure}
\centering \includegraphics[scale= 1]{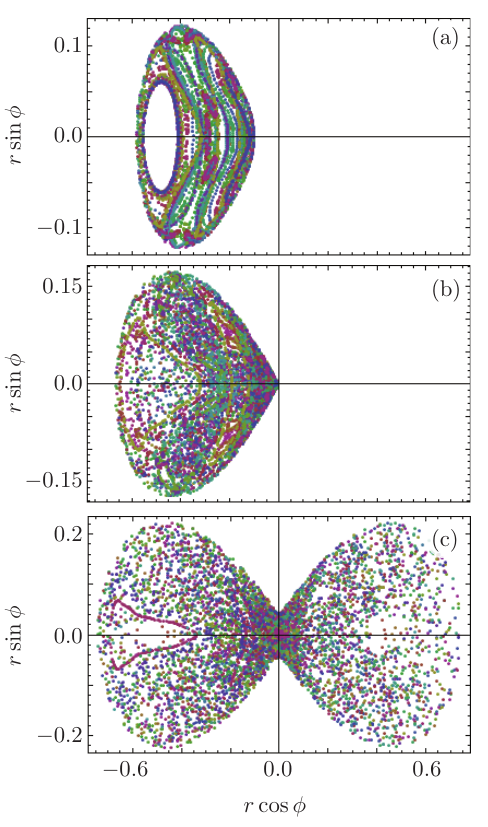}
\caption{(Color online). Poincar\'e sections for phase space $\{ r , \phi \}$ at $p(t)=0$ with $r=1+ j_{z}/j$ and initial scaled energies  $E/ (\omega j) =$ , (a) $-0.99$ (b) $-0.951885$, and (c) $-0.91$. The parameter values used here are: $j=50$, $\omega = \omega_{f}$, $\eta = 0.1 \omega_{f}$, $\lambda = 1.25 \lambda_{c}^{(SC)}$. These deliver a minimum scaled angular momentum projection, $\langle \hat{J}_{z}\rangle / (\omega j) = -0.828784$, with associated scaled energy $E/ (\omega j) = -0.951885$ in the quantum case.   } \label{fig:Figure5}
\end{figure}
%%%%%%%%%%%%%%%%%%%%%%%%%%%%%%%%%%%%%%%%%%%%%%%%%%%%%%%%%%%%%%%%%%%%%%%%%%%%%%%%%%%%%%%%

\begin{figure}
\centering \includegraphics[scale= 1]{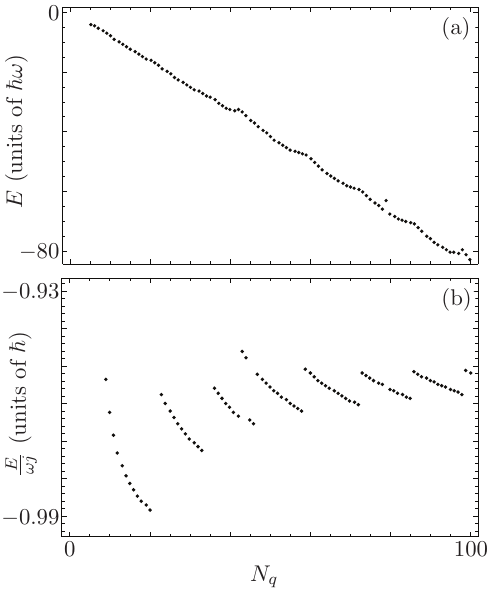}
\caption{(Color online). (a) Energy and (b) scaled energy of the quantum state that delivers a minimum scaled angular momentum projection $\langle \hat{J}_{z}\rangle / (\omega j)$ for the parameters $\omega = \omega_{f}$, $\eta = 0.1 \omega_{f}$, $\lambda = 1.25 \lambda_{c}^{(SC)}$ and ensemble size $N_{q} \in \left[5,100\right]$. } \label{fig:Figure6}
\end{figure}

It was recently shown that for non-interacting qubits there exists a transition from order to chaos for low energy initial states \cite{BastarracheaMagnani2014p032102}.
Here we want to show that the addition of interaction between qubits conserves this behavior.
But, more important, we want to point that, for a large ensemble, the transition from order to disorder near the ground state of the classical system is related to the energy at which the angular momentum projection $j_{z}$ is minimum.
Sadly, we are not able to provide an analytic value for this parameter just numerical examples.
Figure \ref{fig:Figure5} shows Poincar\'e sections at $p(t)=0$ for a system with parameters $j=50$, $\omega = \omega_{f}$, $\eta= 0.1 \omega_{f}$ and $\lambda = 1.25 \lambda_{c}^{(SC)}$ and forty random initial conditions with energy below, Fig. \ref{fig:Figure5}(a), at, Fig. \ref{fig:Figure5}(b), and above, Fig. \ref{fig:Figure5}(c), the scaled energy $E/(\omega j) =  -0.951885$.
The scaled energy was calculated in the quantum model for a maximum number of 125 photons in the field.
Note that below this critical scaled energy the system presents only stable orbits and the allowed phase state is restricted, Fig. \ref{fig:Figure5}(a). 
Then, at the critical scaled energy, Fig. \ref{fig:Figure5}(b), there exists both stable and chaotic orbits and more of the phase space area is available. 
Finally, as the initial state energy becomes larger than the critical energy, more phase space is available and the stable orbits start to diminish in number, Fig. \ref{fig:Figure5}(c).
This behavior was found in a sampling of different parameter sets above the semi-classical critical coupling, $\lambda_{c}^{(SC)}$.
Now, this classical scaled energy corresponds to the that of the quantum state with the minimum scaled value of $\langle \hat{J}_{z} \rangle / (\omega j)$.
Such behavior was confirmed numerically for different ensemble sizes and system parameters. 
For example, for the parameters mentioned above and varying the ensemble size, the quantum state that gives the minimum value of $\langle \hat{J}_{z} \rangle$  has an energy that almost varies linearly in the ensemble size, $N_{q}$, as seen in Fig. \ref{fig:Figure6}.
This was also confirmed for a sampling of different parameter sets slightly above the semi-classical critical coupling, $\lambda_{c}^{(SC)}$.

%%%%%%%%%%%%%%%%%%%%%%%%%%%%%%%%%%%%%%%%%%%%%%%%%%%%%%%%%%%%%%%%%%%%%%%%%%%%%%%%%%%%%%%%
%%%%%%%%%%%%%%%%%%%%%%%%%%%%%%%%%%%%%%%%%%%%%%%%%%%%%%%%%%%%%%%%%%%%%%%%%%%%%%%%%%%%%%%%

%%%%%%%%%%%%%%%%%%%%%%%%%%%%%%%%%%%%%%%%%%%%%%%%%%%%%%%%%%%%%%%%%%%%%%%%%%%%%%%%%%%%%%%%
\section{Conclusions}
%%%%%%%%%%%%%%%%%%%%%%%%%%%%%%%%%%%%%%%%%%%%%%%%%%%%%%%%%%%%%%%%%%%%%%%%%%%%%%%%%%%%%%%%

We studied the ground state of a finite ensemble of interacting qubits driven by a quantum field. 
We found a specific parameter region that delivers a maximally entangled W-state in the ground state of the ensemble.
This parameter region corresponds to the first of a finite series of quantum phase transition in the ground state.
As the ensemble size increases, the area of this first non-vacuum ground state decreases and, in the classical limit when the size of the ensemble is infinite, becomes the critical parameter defining the phase transition of the corresponding classical system. 

In addition, while study the classical analog of the model to find the phase transition, we find a critical energy at which there is a transition from order to disorder in the dynamics of the system. 
We studied numerically the behavior of different parameter sets and found that, in all cases studied, this critical energy is related to the energy of the quantum state that delivers the minimum value of $\langle \hat{J}_{z} \rangle$ for each parameter set.
This transition, which is related to an excited state, is interesting because it occurs for values close to the ground state and not far from it as one would expect.
We plan to extend our research regarding excited state phase transitions of this model in the near future.

%%%%%%%%%%%%%%%%%%%%%%%%%%%%%%%%%%%%%%%%%%%%%%%%%%%%%%%%%%%%%%%%%%%%%%%%%%%%%%%%%%%%%%%%
\section*{Acknowledgments}
%%%%%%%%%%%%%%%%%%%%%%%%%%%%%%%%%%%%%%%%%%%%%%%%%%%%%%%%%%%%%%%%%%%%%%%%%%%%%%%%%%%%%%%%
R.~A.~Robles~Robles acknowledges financial support from CONACYT through the master degree scholarship $\#$298009. 

\bibliographystyle{apsrev}
%\bibliography{D:/ExternalHD/Bibliography/references}

\end{document}